\documentclass[12pt,letter]{article}

\usepackage{graphicx, epsfig, color}
\usepackage{amsmath,amssymb,hyperref}
\usepackage{subfigure}

\def\eslt{\not\!\!{E_T}}
\def\to{\rightarrow}

\def\bi{\begin{itemize}}
\def\ei{\end{itemize}}
\def\te{\tilde e}
\def\ta{\tilde a}

\def\tu{\tilde u}
\def\sps1ap{SPS1a$^\prime$}
\def\c1p{C1$^\prime$}

\def\tb{\tilde b}
\def\tf{\tilde f}

\def\tst{\tilde t}
\def\ttau{\tilde \tau}

\def\tg{\tilde g}
\def\tnu{\tilde\nu}

\def\tw{\widetilde W}
\def\tz{\widetilde Z}
\def\alt{\lesssim}
\def\agt{\gtrsim}
\def\be{\begin{equation}}  
\def\ee{\end{equation}}  
\def\bea{\begin{eqnarray}}  
\def\eea{\end{eqnarray}}  
\def\beas{\begin{eqnarray*}}  
\def\eeas{\end{eqnarray*}}

\newcommand\plb[3]{{\it Phys.\ Lett.\ }{\bf B #1} (#2) #3}

\newcommand{\hepph}[1]{hep-ph/#1}

\begin{document}
\begin{titlepage}
\begin{flushright}
OUHEP-180129
\end{flushright}

\vspace{0.5cm}
\begin{center}
{\Large \bf Anomaly mediated SUSY breaking model\\  
retrofitted for naturalness
}\\ 
\vspace{1.2cm} \renewcommand{\thefootnote}{\fnsymbol{footnote}}
{\large Howard Baer$^1$\footnote[1]{Email: baer@nhn.ou.edu }, 
Vernon Barger$^2$\footnote[2]{Email: barger@pheno.wisc.edu } and
Dibyashree Sengupta$^1$\footnote[3]{Email: Dibyashree.Sengupta-1@ou.edu }
}\\ 
\vspace{1.2cm} \renewcommand{\thefootnote}{\arabic{footnote}}
{\it 
$^1$Department. of Physics and Astronomy,
University of Oklahoma, Norman, OK 73019, USA \\
}
{\it 
$^2$Department. of Physics,
University of Wisconsin, Madison, WI 53706, USA \\
}

\end{center}

\vspace{0.5cm}
\begin{abstract}
\noindent 
Anomaly-mediated supersymmetry breaking (AMSB) models seem to have become increasingly implausible due to 
1. difficulty in generating a Higgs mass $m_h\sim 125$ GeV, 
2. typically unnatural superparticle spectra characterized by a large superpotential mu term and
3. the possibility of a wino-like lightest SUSY particle (LSP) as 
dark matter now seems to be excluded.
In the present paper we propose some minor modifications to the paradigm model 
which solve these three issues. 
Instead of adding a universal bulk scalar mass to avoid tachyonic sleptons, 
we add distinct Higgs and matter scalar soft masses which then allow for light higgsinos. 
To gain accord with the measured Higgs mass, we also include a bulk trilinear soft term. 
The ensuing natural generalized AMSB (nAMSB) model then
has a set of light higgsinos with mass nearby the weak scale $m(W,Z,h)\sim 100$ GeV 
as required by naturalness while the winos populate the several hundred GeV range
and gluinos and squarks occupy the multi-TeV range. 
For LHC searches, the wino pair production followed by decay to same-sign diboson signature 
channel offers excellent prospects for discovery at high luminosity LHC 
along with higgsino pair production leading to soft dileptons plus jet(s)+MET. 
A linear $e^+e^-$ collider operating above higgsino pair production threshold
should be able to distinguish the AMSB gaugino spectra from unified or mirage unified scenarios.
Dark matter is expected to occur as a higgsino-like WIMP plus axion admixture.

\vspace*{0.8cm}

\end{abstract}

\end{titlepage}

\section{Introduction}
\label{sec:intro}

The discovery of $D$-branes in superstring models in the 1990s\cite{Polchinski:1995mt} ushered in
new avenues for particle physics model building. In the case of supersymmetry 
(SUSY), this was exemplified initially with the advent of models 
where the dominant contribution to soft SUSY breaking Lagrangian parameters originated from violations of the superconformal anomaly, in what became known as
anomaly-mediated SUSY breaking models, or AMSB\cite{RS,amsb}.
The AMSB contributions to soft SUSY breaking terms are always present in 
gravity mediation, but since they occur at loop level, they are usually suppressed compared to tree-level contributions and hence had previously been mostly 
neglected. Randall and Sundrum (RS) constructed an extra-dimensional
scenario where the AMSB soft term contributions were expected to 
be the dominant or nearly dominant terms. 
The initial idea was that the visible sector, 
usually assumed to be the Minimal Supersymmetric Standard Model or MSSM, 
would be located on one three-brane extending through an assumed extra-dimensional spacetime, 
while SUSY breaking would occur on a different brane.
Thus, the SUSY breaking sector was in fact {\it sequestered}, or separated from the
visible sector brane within the extra-dimensional spacetime. This setup
suppressed tree-level SUSY breaking soft terms in the visible sector. 
But since gravity propagates in the bulk, 
the entire extra-dimensional spacetime, the anomaly-mediated contributions 
could dominate the visible sector soft terms. 

The AMSB gaugino masses were calculated to be proportional to the corresponding gauge group beta functions times the gravitino mass
\be
M_i=\frac{\beta_i}{g_i} m_{3/2}
\label{eq:m_inos}
\ee
with $\beta_i=\frac{g_i^3}{16\pi^2}b_i$, $b_i=(6.6,1,-3)$ and $i$ labels the gauge group.
Taking into account the running gauge coupling values at the weak scale, 
then one expects  gaugino masses in the ratio $M_1:M_2:M_3\sim 3.3:1:-9$
so that the {\it winos} are the lightest of the weak scale gauginos. 
This is in contrast to models with unified gaugino masses where the bino occurs as the lightest gaugino. 
The lightest neutral wino was then typically assumed to be the
lightest SUSY particle (LSP) in AMSB with striking consequences for
collider and dark matter signatures\cite{Feng:1999fu,GGW,FM}.

In addition, in AMSB the soft breaking scalar masses were computed to be
\be
m_{\tf}^2=-\frac{1}{4}\left\{ \frac{d\gamma}{dg}\beta_g+\frac{d\gamma}{df}\beta_f\right\}m_{3/2}^2
\label{eq:msq}
\ee
where $\beta_f$ is the beta function for the corresponding superpotential Yukawa coupling and 
anomalous dimension $\gamma=\partial\ln Z/\partial\ln\mu$ with $Z$ the 
wave function renormalization constant and $\mu$ is the running energy scale.
The AMSB contribution to trilinear soft SUSY breaking terms is given by
\be
A_f=\frac{\beta_f}{f}m_{3/2}
\label{eq:Aterms}
\ee
where $f$ is the corresponding Yukawa coupling.

For some assumed value of gravitino mass $m_{3/2}\sim 50-100$ TeV, then
all the AMSB soft terms are comparable to each other 
with values near to the weak scale as required by phenomenology.
An annoyance is that the slepton masses turn out to be tachyonic with negative
mass-squared leading to an electric charge breaking minimum for the
scalar potential. It was suggested by RS\cite{RS} that additional bulk 
contributions to scalar masses, which are comparable to the AMSB contributions,
could be present to alleviate this problem. An assortment of other solutions
to the negative slepton mass problem were also devised\cite{negslepmass}.

To gain concrete phenomenological predictions for AMSB at colliding beam 
and dark matter detection experiments, a {\it minimal} AMSB model (mAMSB) was
devised wherein a common bulk contribution $m_0^2$ was appended to all
AMSB scalar mass-squared values\cite{GGW,FM}. 
Once the weak scale soft terms were determined, then the superpotential 
$\mu$ term was tuned so as to maintain the measured value of the $Z$ boson
mass via the scalar potential minimization conditions. 
Thus, the parameter space of the mAMSB model was given by
\begin{equation}
m_0,\ \ m_{3/2},\ \tan\beta ,\ sign(\mu ) .
\label{eq:mAMSB}
\end{equation}
Expectations for LHC searches within the  mAMSB construct have been
presented in Ref's \cite{hb,allanach}. 
Searches for direct chargino pair production in mAMSB with disappearing
tracks from long-lived but ultimately unstable wino-like 
charginos\cite{Feng:1999fu} have been presented by Atlas\cite{atlas}.

The minimal AMSB model has provided a beautiful and compelling framework
for new physics searches. It has been especially appreciated for
containing solutions to the SUSY flavor problem (since the sfermions
of each generation acquire common masses) and the gravitino problem
(since gravitinos are so heavy that they decay much more quickly than
the TeV-scale gravitinos which are expected in usual SUGRA models).
While wino-like WIMPs are thermally underproduced in the mAMSB model, it was
hypothesized by Moroi and Randall\cite{mr} 
that non-thermal WIMP production from, for
instance, decay of light moduli fields could augment the relic abundance of
dark matter and bring its mass abundance into accord with measured values.

While the mAMSB model is a well-motivated and beautiful 
construct, recently it has suffered several setbacks on the phenomenological 
front. 
\begin{itemize} 
\item The first of these was the discovery of the Higgs boson
at a mass value $m_h\simeq 125$ GeV. In the mAMSB model, the trilinear 
soft terms given by Eq. \ref{eq:Aterms} are generally not large enough
to lift the predicted value of $m_h$ into the $125$ GeV range unless 
sparticle masses are very heavy -- in the vicinity of tens of 
TeV\cite{arbey,bbm,arbey2}. 
Such heavy sparticle masses exacerbate the so-called Little Hierarchy problem
which arises from the growing mass gap between the measured value of the 
weak scale and the sparticle mass scale.
\item The second setback arises from non-observation of sparticles 
at the CERN Large Hadron Collider (LHC). While one solution to this issue 
is to simply posit that the mAMSB sparticles are heavier than 
experimental limits, this also makes the theory increasingly 
unnatural\cite{dew} and hence increasingly implausible.
\item A third setback arose on the dark matter front. In mAMSB, 
where a wino-like WIMP is expected to comprise the dark matter, the model
has come into conflict with new stringent limits from direct 
and indirect dark matter detection experiments. 
Searches for WIMPs at underground noble liquid experiments-- 
which test the spin-independent (SI) direct detection (DD) rate-- 
apparently exclude about half the remaining mAMSB parameter space\cite{wimp}.
Meanwhile, indirect WIMP detection (IDD) searches-- 
via observation of gamma rays arising from WIMP-WIMP annihilation into hadrons 
followed by {\it e.g.} $\pi^0\to \gamma\gamma$ decay-- 
have placed severe limits on wino dark matter. 
The Fermi-LAT/MAGIC collaboration\cite{Ahnen:2016qkx}, 
via a search for gamma rays from dwarf spheroidal galaxies, 
now seems to require $m(wino)\agt 700$ GeV.
Along with this, the HESS experiment\cite{::2016jja}, from 254 hours (10 years) 
of observation of continuum gamma rays arising from the galactic center,
now requires $m(wino)\agt 1200$ GeV. If Sommerfeld enhancement effects
are included in the WIMP-WIMP annihlation rate, then wino-like WIMPs
seem to be excluded over their entire mass 
range\cite{Cohen:2013ama,Fan:2013faa,wimp}. 
At first sight, such limits from IDD dark matter searches would seem to 
exclude models like mAMSB with wino-like WIMP dark matter.\footnote{
A possibility which avoids these constraints consists of mixed wino/axion dark matter\cite{Bae:2015rra}.}
\end{itemize}

To expand upon the fine-tuning/naturalness issue, we here adopt the most 
conservative fine-tuning measure, $\Delta_{\rm EW}$~\cite{ltr,rns}. 
The quantity $\Delta_{\rm EW}$ measures how well the weak scale MSSM Lagrangian parameters match the measured value of the weak scale.
By minimizing the MSSM weak scale scalar potential to determine the Higgs field vevs, one derives the well-known expression relating the $Z$-boson mass to
the SUSY Lagrangian parameters:
\be \frac{m_Z^2}{2} = \frac{m_{H_d}^2 +
\Sigma_d^d -(m_{H_u}^2+\Sigma_u^u)\tan^2\beta}{\tan^2\beta -1} -\mu^2
\simeq  -m_{H_u}^2-\Sigma_u^u(\tst_{1,2})-\mu^2 .
\label{eq:mzs}
\ee 
Here, $\tan\beta =v_u/v_d$ is the ratio of Higgs field
vacuum-expectation-values and the $\Sigma_u^u$ and $\Sigma_d^d$
contain an assortment of radiative corrections, the largest of which
typically arise from the top squarks. Expressions for the $\Sigma_u^u$
and $\Sigma_d^d$ are given in the Appendix of Ref.~\cite{rns}. 
Thus, $\Delta_{\rm EW}$ compares the maximal contribution on the 
right-hand-side (RHS) of Eq. \ref{eq:mzs} to the value of $m_Z^2/2$.
If the RHS terms in Eq.~(\ref{eq:mzs}) are individually
comparable to $m_Z^2/2$, then no unnatural fine-tunings are required to
generate $m_Z=91.2$ GeV.\footnote{Other measures include 
$\Delta_{BG}\equiv max_i|\frac{p_i}{m_Z^2}\frac{\partial m_Z^2}{\partial p_i}|$ 
where $p_i$ are fundamental parameters of the theory\cite{bg}. 
In a theory where all soft terms are {\it interdependent}
(such as AMSB or SUGRA or GMSB) then $\Delta_{BG}$ reduces 
to $\Delta_{\rm EW}$\cite{dew}.
Sometimes $\Delta_{HS}\equiv\delta m_h^2/m_h^2$ is used\cite{HS} where 
$\delta m_h^2\sim -\frac{3f_t^2}{8\pi^2}(m_{Q_3}^2+m_{U_3}^2+A_t^2)\ln\left(\Lambda^2/m_{SUSY}^2\right)$ with $f_t$ the top Yukawa coupling, $\Lambda$ is as high
as $m_{GUT}$ and $m_{SUSY}\sim 1$ TeV. This measure has been oversimplified 
by neglecting the $m_{H_u}^2$ contribution to its own running 
so as not to allow for radiatively driven naturalness, where large
high scale soft terms are driven by radiative corrections to natural values 
at the weak scale\cite{comp3,mt}.
}  
 
The main requirements for low electroweak fine-tuning ($\Delta_{\rm
EW}\alt 30$) \footnote{ The onset of fine-tuning for $\Delta_{\rm EW}\agt
30$ is visually displayed in Ref.~\cite{upper}.} are the following.
\bi
\item $|\mu |\sim 100-300$ GeV~\cite{Chan:1997bi,bbh}
(the lighter the better) 
where $\mu \agt 100$ GeV is required to accommodate LEP2 limits 
from chargino pair production searches.
\item $m_{H_u}^2$ is driven radiatively to small-- not large--
negative values at the weak scale~\cite{ltr,rns}.
\item The top squark contributions to the radiative corrections
$\Sigma_u^u(\tst_{1,2})$ are minimized for TeV-scale highly mixed top
squarks~\cite{ltr}. This latter condition also lifts the Higgs mass to
$m_h\sim 125$ GeV. For $\Delta_{\rm EW}\alt 30$, the lighter top
squarks are bounded by $m_{\tst_1}\alt 3$ TeV~\cite{rns,upper}.
\item The gluino mass, which feeds into the stop masses at one-loop 
and hence into the scalar potential at two-loop order,
is bounded by $m_{\tg}\alt 6$ TeV~\cite{rns,upper}.
\ei

\begin{figure}[tbp]
\begin{center}
\includegraphics[height=0.4\textheight]{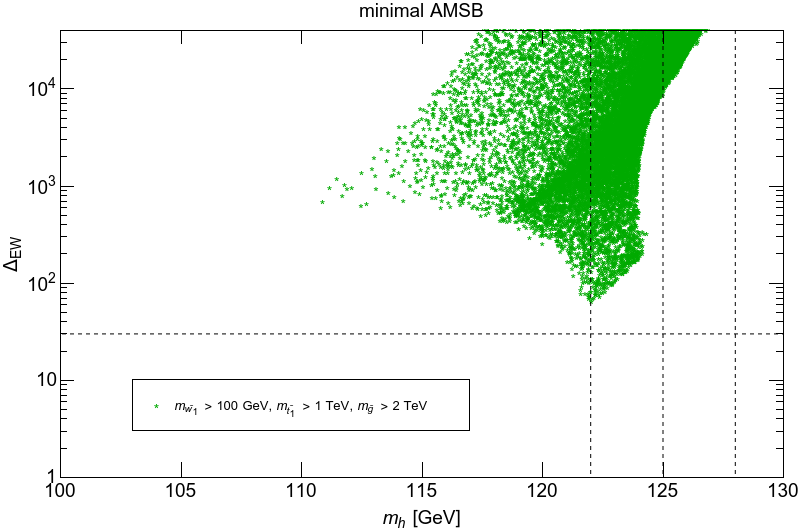}
\caption{Plot of points from a scan over mAMSB parameter space in 
the $\Delta_{\rm EW}$ vs. $m_h$ plane.
\label{fig:dew_mh_1}}
\end{center}
\end{figure}

In Fig. \ref{fig:dew_mh_1}, we show the results of a scan over mAMSB model 
parameter space in the $\Delta_{\rm EW}$ vs. $m_h$ plane.
We use Isajet 7.87\cite{isajet} to generate the mAMSB spectra.
We have scanned over 
\bi
\item $m_0: 1-10$ TeV,
\item $m_{3/2}: 80-1000$ TeV,
\item $\tan\beta : 4 - 58$,
\ei
with $\mu >0$.

From Fig. \ref{fig:dew_mh_1}, we see that the minimal value of 
$\Delta_{\rm EW}$ occurs around 100 so that indeed the model is fine-tuned 
in the electroweak sector at least at the $\sim 1\%$ level. 
The lowest $\Delta_{\rm EW}$ points occur at $m_{3/2}\sim 100$ TeV where
$m_{\tg}\sim 2$ TeV, just beyond the current LHC 
$m_{\tg}$ mass limit\cite{lhc_gl}. 
While many of these points have $m_h\sim 122$ GeV, to gain $m_h\sim 125$ GeV 
the value of $\Delta_{\rm EW}$ jumps to $\agt 6000$.

To improve upon this situation, in this paper 
we present a retrofitted phenomenological AMSB model
which is a generalization of mAMSB and which addresses the three issues 
discussed above.
Indeed, in the original RS paper\cite{RS}, the authors actually advocated for 
the modifications we present here. 
It was only when some simplifications were implemented in the 
original minimal AMSB model that these features were abandoned\cite{GGW,FM}. 
The two generalizations to mAMSB include the following:
\begin{enumerate}
\item independent bulk contributions $m_{H_u}^2(bulk)$ and 
$m_{H_d}^2(bulk)$ to the soft SUSY breaking Higgs masses as 
opposed to matter scalar bulk masses $m_0^2(1,2)$ (for first/second 
generation matter scalars) and $m_0^2(3)$ 
(for third generation matter scalars) and
\item inclusion of bulk contributions $A_0$ to the trilinear soft terms.
\end{enumerate}
These two modest changes in the AMSB model will allow each of the three
issues above to be circumvented. However, we will also see that the 
anticipated collider phenomenology and dark matter expectations will
be very different. 
After bringing the model into accord with the measured Higgs mass and 
naturalness, the LSP  will no longer be a wino-like neutralino, but instead
a higgsino-like neutralino. 
If we posit that the SUSY $\mu$ problem is solved via the Kim-Nilles mechanism\cite{KN} 
(a supersymmetrized version of the DFSZ axion model\cite{dfsz} which allows
for $\mu\ll m_{soft}$) then dark matter is expected to consist of
an axion plus higgsino-like WIMP admixture\cite{bbc}.

In Sec. \ref{sec:nAMSB}, we make explicit our modified AMSB soft term formulae.
We also present aspects of the anticipated natural AMSB spectra where now
the LSP is expected to be a higgsino-like neutralino but where the
lightest gaugino is still expected to be wino-like.
Since the model can now be rendered natural, we dub the resultant model
as nAMSB, or natural anomaly-mediation, to distinguish it from the
previously explored minimal AMSB model.
We present some benchmark spectra and a nAMSB model line.
In Sec. \ref{sec:lhc_dm}, we discuss consequences of the nAMSB model for
collider and dark matter searches.
In Sec. \ref{sec:conclude}, we summarize and present our conclusions.

\section{Natural Anomaly Mediated SUSY Breaking Model (nAMSB)}
\label{sec:nAMSB}

\subsection{Soft terms for nAMSB}

In this Section, we propose several minor modifications of the mAMSB model
which will allow for naturalness along with a Higgs mass $m_h\simeq 125$ GeV.

For gaugino masses, we maintain the usual formulae:
\bea
M_1&=& {33\over 5}{g_1^2\over 16\pi^2}m_{3/2} ,\label{eq:namsb1}\\
M_2&=& {g_2^2\over 16\pi^2}m_{3/2} ,\ \\
M_3&=& -3{g_3^2\over 16\pi^2}m_{3/2} .
\eea

Third generation soft SUSY breaking scalar squared masses are given by
\bea
m_{U_3}^2 &=& \left(-{88\over 25}g_1^4+8g_3^4+2f_t\hat{\beta}_{f_t}\right)
{m_{3/2}^2\over (16\pi^2)^2}+m_0^2(3), \label{eq:amsbU}\\
m_{D_3}^2 &=& \left(-{22\over 25}g_1^4+8g_3^4+2f_b\hat{\beta}_{f_b}\right)
{m_{3/2}^2\over (16\pi^2)^2}+m_0^2(3),\\
m_{Q_3}^2 &=& \left(-{11\over 50}g_1^4-{3\over 2}g_2^4+
8g_3^4+f_t\hat{\beta}_{f_t}+f_b\hat{\beta}_{f_b}\right)
{m_{3/2}^2\over (16\pi^2)^2}+m_0^2(3),\\
m_{L_3}^2 &=& \left(-{99\over 50}g_1^4-{3\over 2}g_2^4+
f_\tau\hat{\beta}_{f_\tau}\right) {m_{3/2}^2\over (16\pi^2)^2}+m_0^2(3),\\
m_{E_3}^2 &=& \left(-{198\over 25}g_1^4+
2f_\tau\hat{\beta}_{f_\tau}\right) {m_{3/2}^2\over (16\pi^2)^2}+m_0^2(3),
\eea
while first/second generation scalar squared masses are given 
by similar formulae but where the associated Yukawa couplings may be safely ignored and the bulk sfermion mass is changed from $m_0^2(3)\to m_0^2(1,2)$.

For soft SUSY breaking Higgs masses, we propose (in accord with Ref. \cite{RS})
that each Higgs doublet receive an independent bulk mass contribution so that
\bea
m_{H_u}^2 &=& \left(-{99\over 50}g_1^4-{3\over 2}g_2^4+
3f_t\hat{\beta}_{f_t}\right) {m_{3/2}^2\over (16\pi^2)^2}+m_{H_u}^2(bulk),\\
m_{H_d}^2 &=& \left(-{99\over 50}g_1^4-{3\over 2}g_2^4+
3f_b\hat{\beta}_{f_b}+f_\tau\hat{\beta}_{f_\tau}\right) 
{m_{3/2}^2\over (16\pi^2)^2}+m_{H_d}^2(bulk) .\label{eq:amsbHd}
\eea
The freedom of independent bulk Higgs soft masses $m_{H_u}^2(bulk)$ and 
$m_{H_d}^2(bulk)$
may be traded using the electroweak minimation conditions for the alternative 
{\it weak scale} inputs $\mu$ and $m_A$ (as in the NUHM2 SUSY model\cite{nuhm2}).

\begin{figure}[tbp]
\begin{center}
\includegraphics[height=0.4\textheight]{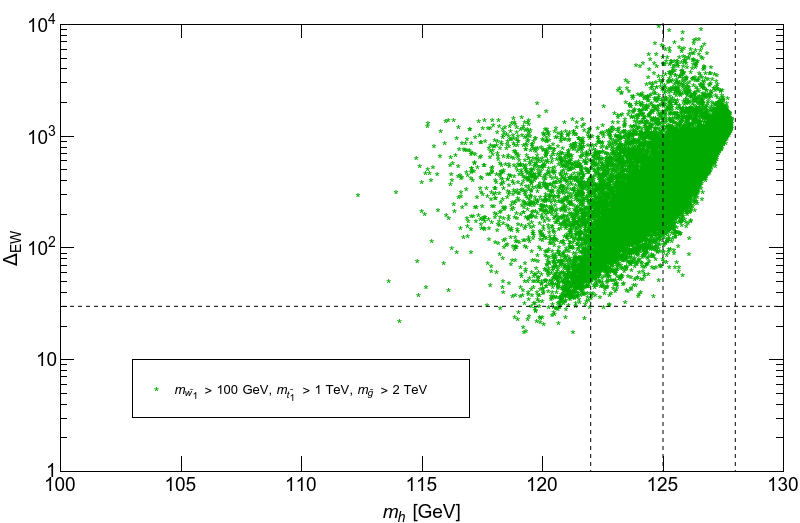}
\caption{Plot of points in the $\Delta_{\rm EW}$ vs. $m_h$ plane 
from a scan over AMSB parameter space 
with added bulk Higgs soft terms but without bulk $A_0$ terms. 
\label{fig:dew_mh_2}}
\end{center}
\end{figure}

Using this flexibility, we again scan over AMSB parameters 
as in Sec.~\ref{sec:intro} but now also including
\bi
\item $\mu :\ 100-500$ GeV and 
\item $m_A:\ 0.25-10$ TeV.
\ei
The results are plotted again in the $\Delta_{\rm EW}$ vs. $m_h$ 
plane and shown in Fig. \ref{fig:dew_mh_2}.
From the figure, we see that now many points have dropped into the natural
area where $\Delta_{\rm EW}<30$. However, almost all these points also have
$m_h\alt 122$ GeV. 

Thus, following Ref. \cite{RS}, we propose adding as well a bulk contribution 
to the trilinear soft terms.
Then the $A$-parameters are given by
\bea
A_t&=&{\hat{\beta}_{f_t}\over f_t}{m_{3/2}\over 16\pi^2}+A_0,\\
A_b&=&{\hat{\beta}_{f_b}\over f_b}{m_{3/2}\over 16\pi^2}+A_0,\ {\rm and} \\
A_\tau &=&{\hat{\beta}_{f_\tau}\over f_\tau}{m_{3/2}\over 16\pi^2}+A_0.
\label{eq:namsbA}
\eea
The quantities $\hat{\beta}_{f_i}$ that enter the expressions for scalar
masses and $A$-parameters are given by the standard expressions
\bea
\hat{\beta}_{f_t} &=& 16\pi^2\beta_t=f_t\left( -{13\over 15}g_1^2-3g_2^2
-{16\over 3}g_3^2+6f_t^2+f_b^2\right),\\
\hat{\beta}_{f_b} &=& 16\pi^2\beta_b 
= f_b\left( -{7\over 15}g_1^2-3g_2^2
-{16\over 3}g_3^2+f_t^2+6f_b^2+f_\tau^2\right),\\
\hat{\beta}_{f_\tau} &=& 16\pi^2\beta_\tau=f_\tau
\left( -{9\over 5}g_1^2-3g_2^2+3f_b^2+4f_\tau^2\right).
\eea

The first two generations of squark and slepton masses are given by the 
corresponding formulae above with the Yukawa couplings set to zero.
Eq.~(\ref{eq:namsb1})-(\ref{eq:namsbA}) serve as RGE boundary conditions
at $Q=m_{\rm GUT}$. 
The nAMSB model is therefore characterized by the parameter set,
\be
m_0(1,2),\ m_0(3),\ m_{3/2},\ A_0,\ \tan\beta ,\ \mu ,\ {\rm and}\ m_A .
\label{eq:pspace}
\ee

\begin{figure}[t]
  \centering
  {\includegraphics[width=.48\textwidth]{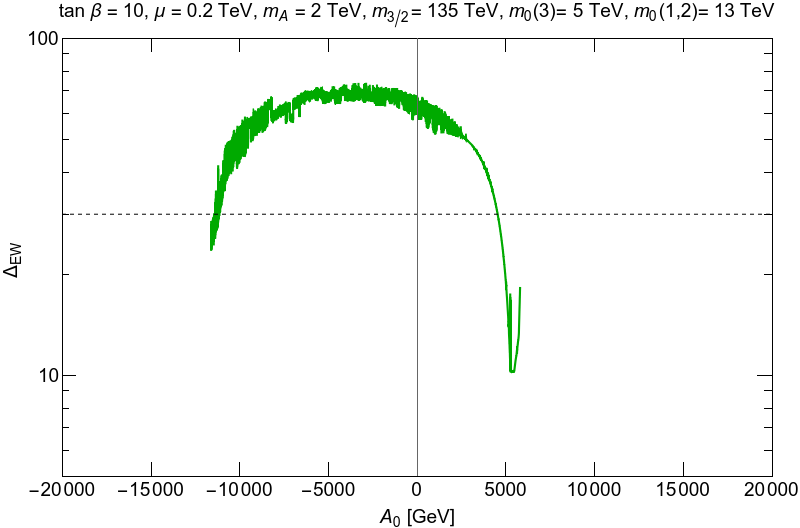}}\quad
  {\includegraphics[width=.48\textwidth]{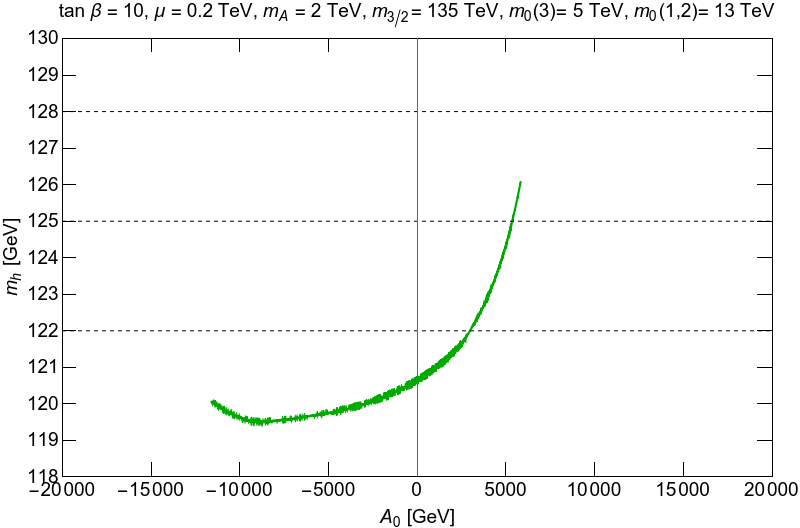}}
  \caption{Frame {\it a}): $\Delta_{\rm EW}$ vs. $A_0$ for
$m_{3/2}=135$ TeV, $m_0(1,2)=13.5$ TeV, $m_0(3)=5$ TeV, $\mu=200$ GeV and
$m_A=2000$ GeV.
In frame {\it b}), we plot $m_h$ vs. $A_0$ for the same parameters.
}  
\label{fig:a0}
\end{figure}

To see the effect of including the bulk $A_0$ trilinear soft term, 
we adopt a nAMSB benchmark point with parameters $m_{3/2}=135$ TeV, 
$m_0(1,2)=13$ TeV, $m_0(3)=5$ TeV, $\mu =200$ GeV and $m_A=2$ TeV 
with $\tan\beta =10$. 
In Fig. \ref{fig:a0}{\it a}), we show the value of $\Delta_{\rm EW}$ as 
we vary $A_0$. For no bulk trilinear, with $A_0=0$, then $\Delta_{\rm EW}\sim 70$
and the model requires EW fine-tuning at the 1.4\% level. As $A_0$ varies
and becomes large positive or negative, then large mixing in the stop sector
leads to a reduction in both $\Sigma_u^u(\tst_{1,2})$ values.
For $A_0\sim +5$ TeV, then $\Delta_{\rm EW}$ drops to as low as 10.
In frame {\it b}), we show the variation in $m_h$ versus $A_0$. With no
bulk contribution to $A$ terms, then $m_h\sim 120$ GeV. 
As $A_0$ increases to $\sim +5$ TeV, then the added stop mixing 
increases $m_h$ until it reaches the $\sim 125$ GeV level. 

\begin{figure}[tbp]
\begin{center}
\includegraphics[height=0.4\textheight]{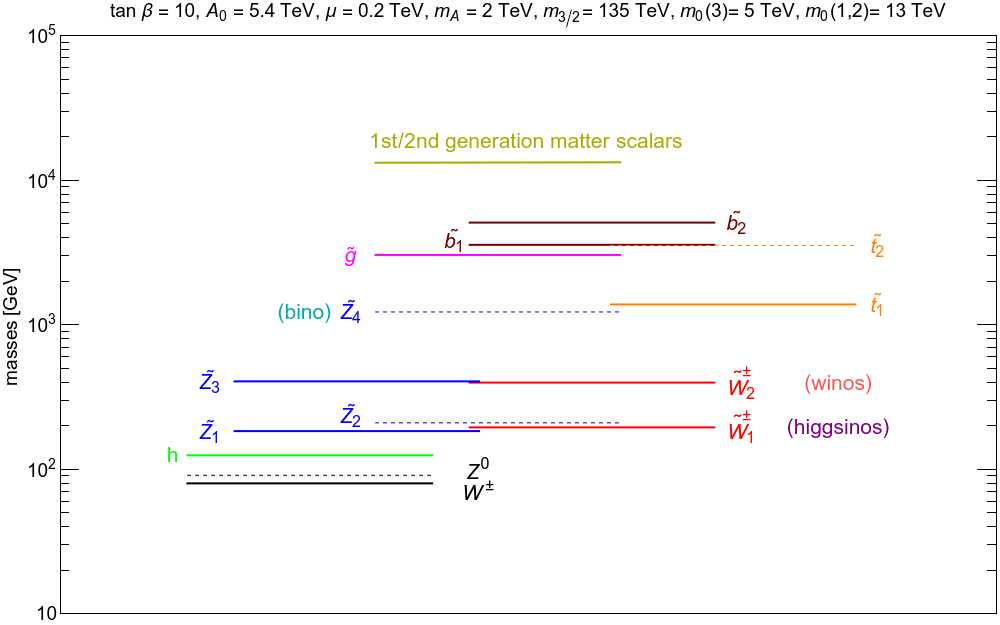}
\caption{A typical superparticle mass spectrum generated from 
natural generalized anomal mediation (nAMSB) as in Table \ref{tab:bm}.
\label{fig:spect}}
\end{center}
\end{figure}

In Fig. \ref{fig:spect}, we show the nAMSB spectra plot from our benchmark 
point where now we adopt $A_0=+5.4$ TeV. 
From the plot, we see that the $W$, $Z$ and $h$ are clustered around the 
$\sim 100$ GeV scale with the higgsinos $\tw_1^\pm$ and $\tz_{1,2}$ 
clustered not too far away at $\sim 200$ GeV as required by naturalness. 
Meanwhile, first/second generation matter sfermions lie in the multi-TeV range at $\sim 13$ TeV. 
For the gauginos, we have $m_{\tg}\sim 3$ TeV, 
well beyond current LHC limits which at present require $m_{\tg}\agt 2$ TeV.\footnote{
This spectra is rather similar to that expected by Dine from the intermediate branch of the 
IIB string theory landscape\cite{Dine:2005iw}.} 
What is characteristic about nAMSB is the rather 
light winos $\tw_2^\pm$ and $\tz_3$ with mass $\sim 400$ GeV. 
The bino $\tz_4$ has mass $\sim 1.2$ TeV. 
For the top squarks, we find them to be highly mixed by the large $A_t$ term 
with $m_{\tst_1}\sim 1.4$ TeV and $m_{\tst_2}\sim 3.5$ TeV. 
As we shall see, the nAMSB mass spectrum leads to very different expectations 
for LHC signatures as compared to the old mAMSB model.

The precise benchmark point mass values are listed numerically in Table \ref{tab:bm}
along with various calculated dark matter and $B$-decay observables.
For this point, the thermal WIMP abundance of higgsino-like WIMP comes in
(from IsaReD\cite{isared}) at $\Omega_{\tz_1}^{TP}h^2\sim 0.009$, 
a factor 13.3 below the measured abundance. 
In the case of nAMSB, we also expect the presence of a
SUSY-DFSZ axion which would likely make up the remaining dark matter abundance.
A complete calculation requires an eight-coupled Boltzmann equation 
computation\cite{bbls}.
The WIMP detection rates are also given, but in this case they must be scaled down
by factors of $\xi\equiv \Omega_{\tz_1}h^2/0.12$ for the SI and SD 
direct detection rates. For the IDD detection rate, the higgsino-like WIMPs
mainly annihilate into the $WW$ channel but in this case must be scaled down 
by a factor $\xi^2$. These rescalings, due to diminished WIMP number density, 
bring the detection rates near or below current experimental limits
(see Ref. \cite{wimp} for a recent summary). 
The naturalness parameter for the benchmark point lies at $\Delta_{\rm EW}=10.2$ 
so the model is quite natural with just $\sim 10\%$ EW fine-tuning required.
%
\begin{table}\centering
\begin{tabular}{lc}
\hline
parameter & nAMSB1 \\
\hline
$m_{3/2}$      & 135000  \\
$\tan\beta$    & 10  \\
$m_0(1,2)$      & 13000 \\
$m_0(3)$      & 5000 \\
$A_0$      & 5400 \\
\hline
$\mu$          & 200   \\
$m_A$          & 2000  \\
\hline
$m_{\tg}$   & 3037.6  \\
$m_{\tu_L}$ & 13189.1  \\
$m_{\tu_R}$ & 13280.1  \\
$m_{\te_R}$ & 12909.7  \\
$m_{\tst_1}$& 1380.8  \\
$m_{\tst_2}$& 3536.0  \\
$m_{\tb_1}$ & 3569.5  \\
$m_{\tb_2}$ & 5085.0  \\
$m_{\ttau_1}$ & 4670.6  \\
$m_{\ttau_2}$ & 4930.8  \\
$m_{\tnu_{\tau}}$ & 4903.1  \\
$m_{\tw_2}$ & 398.0  \\
$m_{\tw_1}$ & 195.1  \\
$m_{\tz_4}$ & 1225.6  \\ 
$m_{\tz_3}$ & 405.9  \\ 
$m_{\tz_2}$ & 209.7  \\ 
$m_{\tz_1}$ & 183.9  \\ 
$m_h$       & 125.1  \\ 
\hline
$\Omega_{\tz_1}^{std}h^2$ & 0.009  \\
$BF(b\to s\gamma)\times 10^4$ & $3.2$  \\
$BF(B_s\to \mu^+\mu^-)\times 10^9$ & $3.8$ \\
$\sigma^{SI}(\tz_1, p)$ (pb) & $9.8\times 10^{-9}$  \\
$\sigma^{SD}(\tz_1 p)$ (pb) & $2.4\times 10^{-4}$  \\
$\langle\sigma v\rangle |_{v\to 0}$  (cm$^3$/sec)  & $2.8\times 10^{-25}$ \\
$\Delta_{\rm EW}$ & 10.2 \\
\hline
\end{tabular}
\caption{Input parameters and masses in~GeV units
for a natural generalized anomaly mediation SUSY benchmark point
with $m_t=173.2$ GeV.
}
\label{tab:bm}
\end{table}

\begin{figure}[tbp]
\begin{center}
\includegraphics[height=0.4\textheight]{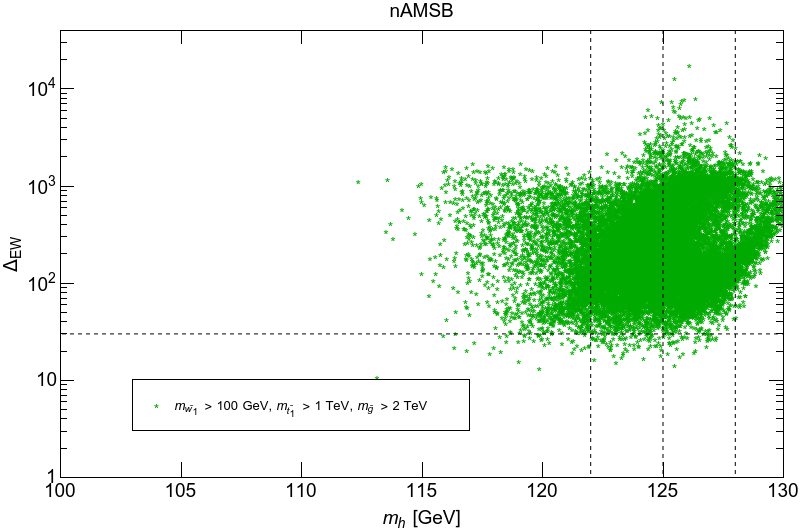}
\caption{Plot of points from a scan over nAMSB parameter space in 
the $\Delta_{\rm EW}$ vs. $m_h$ plane.
\label{fig:dew_mh_3}}
\end{center}
\end{figure}

In Fig. \ref{fig:dew_mh_3}, we repeat the above AMSB parameter space scans
except now we include as well a scan over
\bi
\item $A_0: -20\ \to +20$ TeV. 
\ei
From the figure, we now see data points in accord with LHC sparticle mass constraints which populate the $\Delta_{\rm EW}<30$ naturalness regime whilst 
also allowing for $m_h\sim 125\pm 3$ GeV. 
Thus, the combination of independent bulk Higgs masses and an added bulk
trilinear soft term $A_0$ allows us to bring the AMSB model into accord 
with LHC Higgs mass measurements and naturalness requirements and
dark matter constraints.

\subsection{A nAMSB model line}
\label{ssec:mline}

In phenomenological studies of models for new physics, it is frequently 
useful to adopt {\it model lines} wherein new particle masses increase 
in a controlled manner thus allowing for collider reach 
calculations\cite{mline}, decoupling, etc.
We may elevate our previous benchmark model to a model line by allowing the
gravitino mass to float so all sparticle masses increase with $m_{3/2}$ from
the LHC limits until they become unnatural or decouple.

\begin{figure}[tbp]
\begin{center}
\includegraphics[height=0.3\textheight]{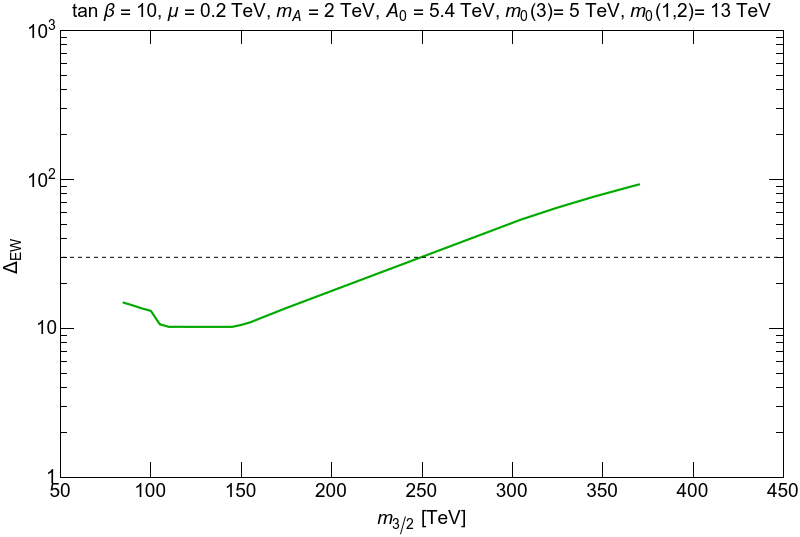}\\
\includegraphics[height=0.3\textheight]{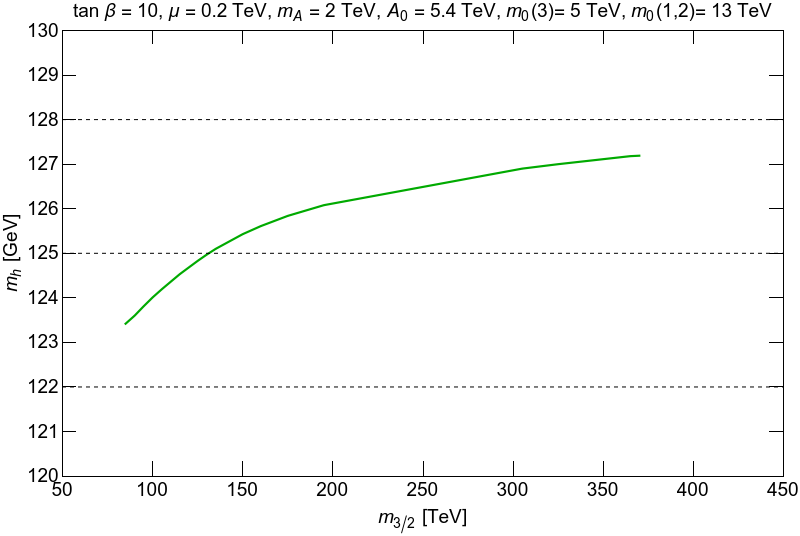}\\
\includegraphics[height=0.3\textheight]{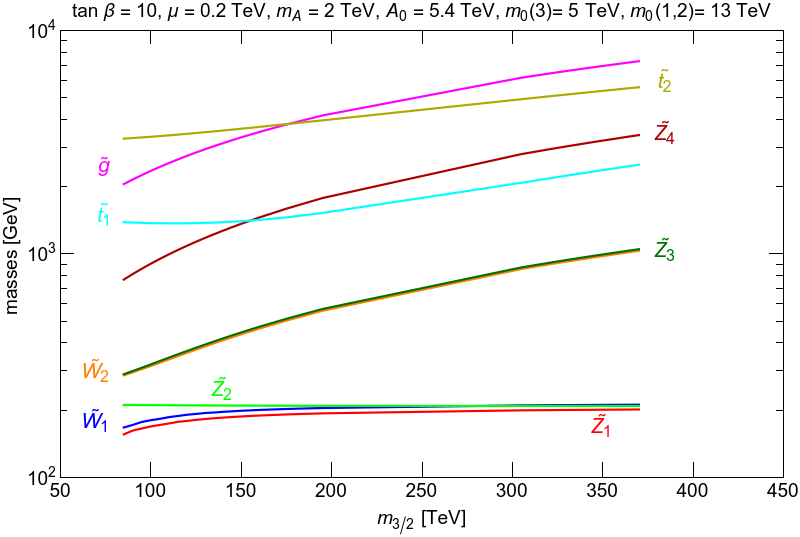}
\caption{Plot of {\it a}) $\Delta_{\rm EW}$, {\it b}) $m_h$ and
{\it c}) various sparticle masses versus $m_{3/2}$ along a nAMSB model line with
$m_0(1,2)=13$ TeV, $m_0(3)=5$ TeV, $A_0=5.4$ TeV, $m_A=2$ TeV and
$\mu=200$ GeV with $\tan\beta =10$.
\label{fig:mline}}
\end{center}
\end{figure}

In Fig. \ref{fig:mline}, we show three frames resulting from a nAMSB 
model line versus $m_{3/2}$ starting at $m_{3/2}\simeq 80$ TeV. 
This latter value corresponds to $m_{\tg}\sim 2$ TeV, 
just beyond the current LHC limits from simplified models\cite{lhc_gl}.
In frame {\it a}), we show how $\Delta_{\rm EW}$ varies. At lower values 
$m_{3/2}\sim 100-150$ TeV, then $\Delta_{\rm EW}\sim 10$ and the model is highly
natural. As $m_{3/2}$ increases, all soft terms increase 
according to Equations~\ref{eq:namsb1}-\ref{eq:namsbA}. 
As $m_{3/2}$ increases to the vicinity of 
250 TeV, then $\Delta_{\rm EW}$ has moved beyond the 30 value where
fine-tuning begins to be required in the weak scale scalar potential. 
Thus, the regime 
where $m_{3/2}\alt 250$ TeV seems favored from a naturalness perspective.
In frame {\it b}), we show the corresponding value of $m_h$ along the
nAMSB model line. 
Its value begins at $m_h\sim 124$ GeV for $m_{3/2}\sim 80$ TeV and increases to $\sim 127$ GeV for $m_{3/2}$ as high as 370 TeV. Thus, the
light Higgs mass stays within its required range (allowing for $\sim \pm 2$ GeV
theory error in our $m_h$ calculation) over the entire model line.
In frame {\it c}), we show various sparticle masses along the model line.
The higgsinos $\tw_1^\pm$ and $\tz_{1,2}$ remain clustered at $\sim 200$ GeV
since the $\mu$  parameter remains fixed. The gluinos and stops lie in the
several TeV range and as their masses increase, so too do the 
radiative corrections $\Sigma_u^u(\tst_{1,2})$ in Eq. \ref{eq:mzs}.
Over the range of $m_{3/2}$ consistent with naturalness, 
$m_{\tg}$ varies from $2-4$ TeV while the lighter stop ranges from
$m_{\tst_1}\sim 1.3-1.5$ TeV. Of considerable interest for collider searches 
is the range of the wino masses $m_{\tw_2}^\pm$ and $m_{\tz_3}$. These vary from
300 GeV for $m_{3/2}\sim 100$ TeV to $\sim 600$ GeV for $m_{3/2}\sim 250$ GeV.
This will have important ramifications for discussion of collider searches
in the next Section.

\subsection{Locus of natural AMSB parameters}
\label{ssec:params}

It is important to check from scans over the full 
generalized AMSB parameter space in Eq.~\ref{eq:pspace} 
where exactly the natural solutions with low $\Delta_{\rm EW}$ exist.
Thus, here we implement a scan over the full parameter 
space and plot each parameter versus $\Delta_{\rm EW}$.
To aid the reader, we show the demarcation where
$\Delta_{\rm EW}$ exceeds 30, although it is simple to extract 
parameter locales for other choices of a maximal $\Delta_{\rm EW}$ value.

\begin{figure}[tbp]
\begin{center}
\includegraphics[height=0.25\textheight]{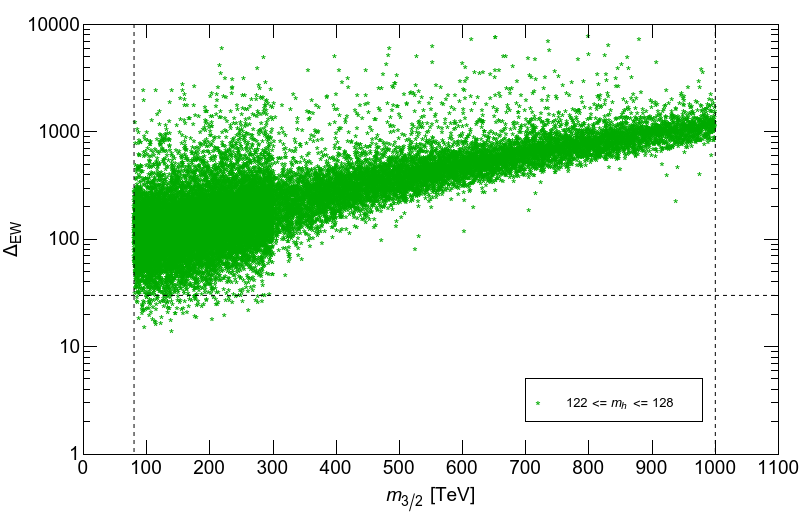}
\includegraphics[height=0.25\textheight]{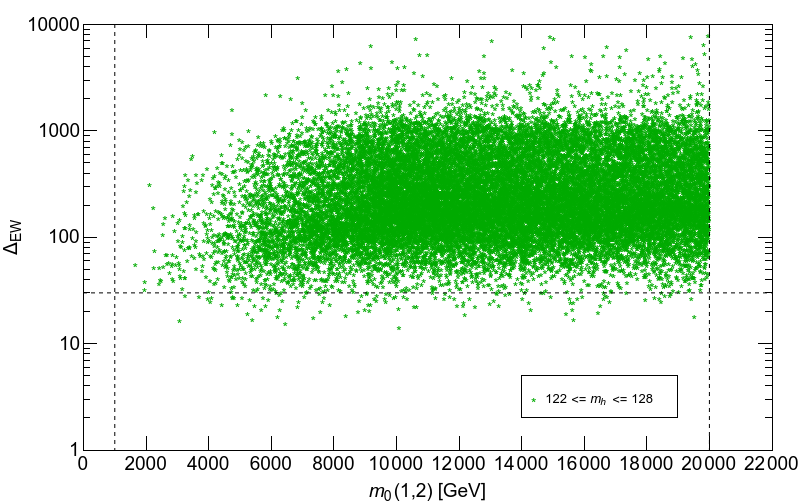}
\includegraphics[height=0.25\textheight]{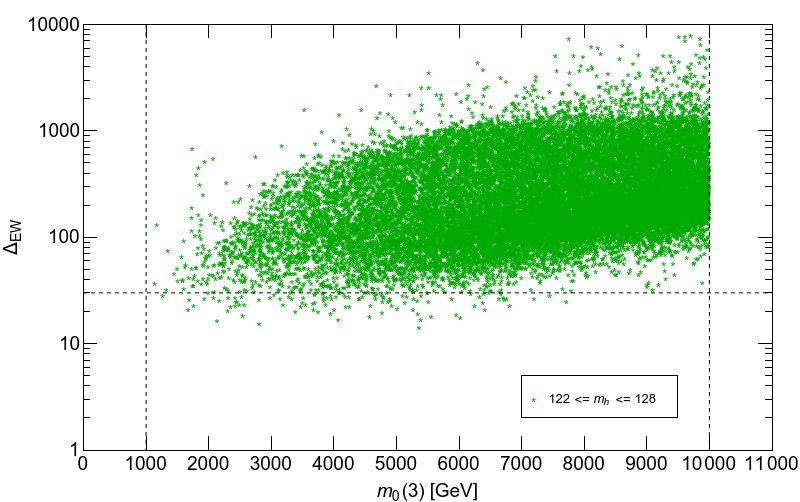}
\caption{Plot of nAMSB parameter scan in the $\Delta_{\rm EW}$ vs. 
{\it a}) $m_{3/2}$, {\it b}) $m_0(1,2)$ and $m_0(3)$ planes.
The greater density of points for $m_{3/2}\alt 300$ TeV comes from the
narrow scan added to the broad scan.
\label{fig:param1}}
\end{center}
\end{figure}

In Fig. \ref{fig:param1}{\it a}), 
we show $\Delta_{\rm EW}$ versus $m_{3/2}$ from our scan.
Our points are extracted from the general scan with limits given above 
and also from a dedicated scan over parameters where $\Delta_{\rm EW}$ 
is more likely to be $\alt 30$: 
$m_{3/2}\sim 80-300$ GeV, $\mu: 100-350$ GeV and $A_0:0.5 m_0(3)- 2m_0(3)$.
All points have $122$ GeV$<m_h<128$ GeV.
From frame {\it a}), we see that to maintain naturalness, 
$m_{3/2}$ is roughly bounded from above by about 300 GeV 
(in accord with the above nAMSB model line). 

In frame {\it b}), we show $\Delta_{\rm EW}$ versus $m_0(1,2)$. 
The first and second generation scalar masses enter the naturalness measure
via electroweak $D$-term contributions\cite{rns,deg} and these terms
tend to cancel for nearly degenerate matter scalars. 
Thus, a wide range of $m_0(1,2)$ values extending up into the 10-20 TeV range 
are allowed by naturalness. Such large first/second generation 
matter scalar masses allow for at least a partial decoupling
solution to the SUSY flavor and CP problem (which may re-arise with the addition of flavor dependent bulk soft terms).

In Fig. \ref{fig:param1}{\it c}) we show $\Delta_{\rm EW}$ vs. $m_0(3)$.
In this case, an upper bound of $m_0(3)\alt 8$ TeV emerges. This is because
for too large values of third generation matter scalars, then 
the $\Sigma_u^u(\tst_{1,2})$ contributions become large thus requiring some 
electroweak fine-tuning.

\begin{figure}[tbp]
\begin{center}
\includegraphics[height=0.22\textheight]{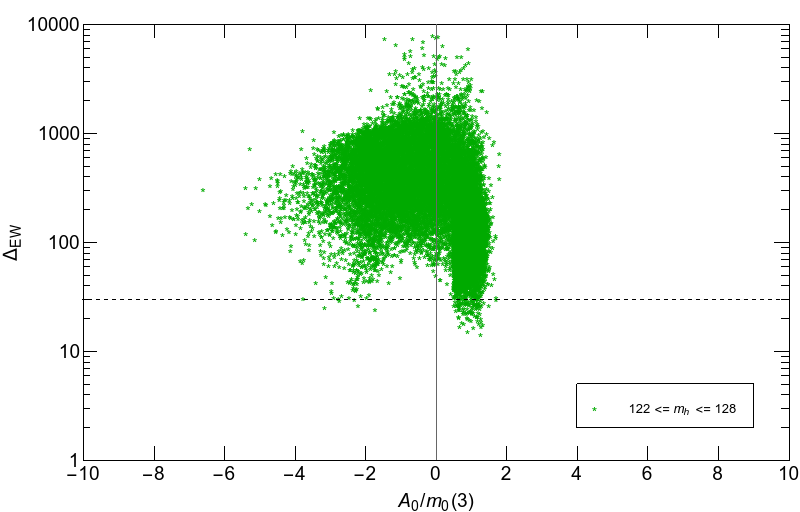}
\includegraphics[height=0.22\textheight]{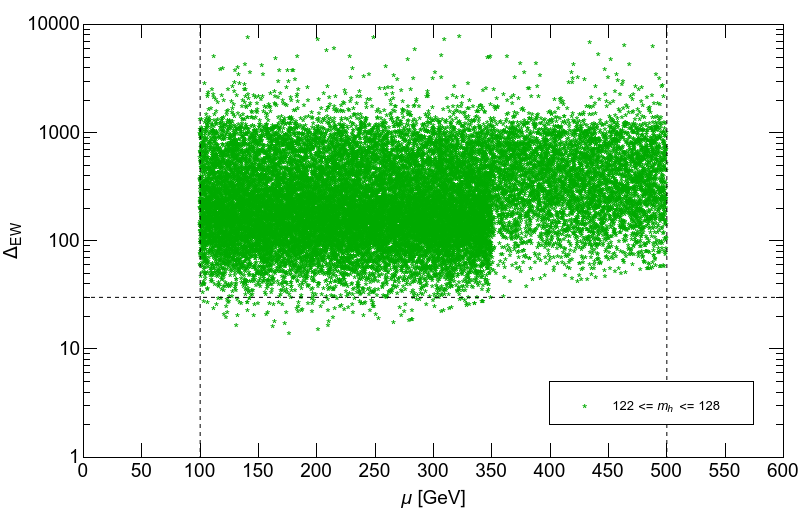}
\includegraphics[height=0.22\textheight]{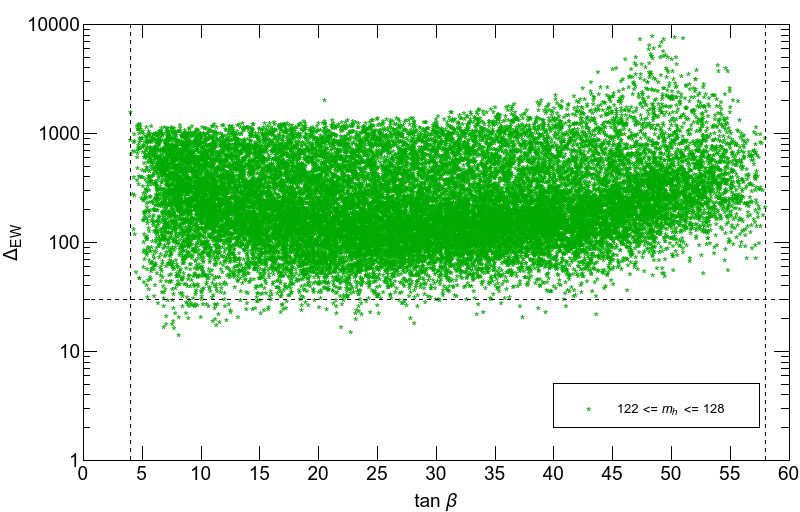}
\includegraphics[height=0.22\textheight]{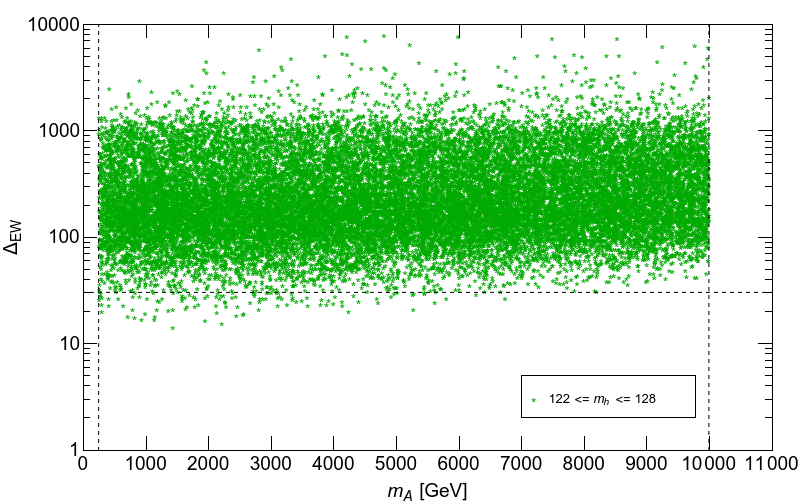}
\caption{Plot of nAMSB parameter scan in the $\Delta_{\rm EW}$ vs. 
{\it a}) $A_0/m_0(3)$, {\it b}) $\mu$, {\it c}) $\tan\beta$ and {\it d}) $m_A$ 
planes.
\label{fig:param2}}
\end{center}
\end{figure}

In Fig. \ref{fig:param2}{\it a}), we show $\Delta_{\rm EW}$ vs. $A_0/m_0(3)$.
Here we see that for $A_0\sim 0$, then $\Delta_{\rm EW}$ is always $\agt 30$
and unnatural. For $A_0/m_0(3)\sim-2$, then $\Delta_{\rm EW}$ drops
below 30. This occurs even more sharply for $A_0/m_0(3)\sim +1$. 
As noted previously, the large $A_0$ values decrease the 
$\Sigma_u^u(\tst_{1,2})$ contributions whilst lifting 
$m_h\sim 125$ GeV\cite{ltr}.
Frame {\it b}) shows $\mu$ vs. $\Delta_{\rm EW}$. Here we see a sharp 
demarcation for naturalness when $\mu \alt 350$ GeV, the lighter the better.
This is also seen from direct computation from Eq. \ref{eq:mzs}.
In frame {\it c}), we show variation versus $\tan\beta$. 
In this case, a wide range of $\tan\beta$ is allowed by naturalness, but not
the very highest values where $\tan\beta\agt 40$. For such high $\tan\beta$, 
then the $\Sigma_u^u(\tb_{1,2})$ may become large, thus requiring 
some fine-tuning. In frame {\it d}), we show variation with $m_A$. 
In this case, for $m_A\gg m_Z$, then $m_{H_d}\sim m_A$ and 
naturalness in Eq. \ref{eq:mzs} would require 
$m_A/\tan\beta\alt\sqrt{30(m_Z^2/2)}$. 
This requires $m_A$ to be bounded from above by about 7-8 TeV.

\subsection{Bounds on sparticle masses in the natural AMSB model}
\label{ssec:mass}

It is desirable in any SUSY model to extract upper bounds on various
sparticle masses from naturalness in order to establish a
{\it testability} criterion for the model. Thus, in this Section, 
we implement the full scan over nAMSB parameter space (as delineated above).

\begin{figure}[tbp]
\begin{center}
\includegraphics[height=0.4\textheight]{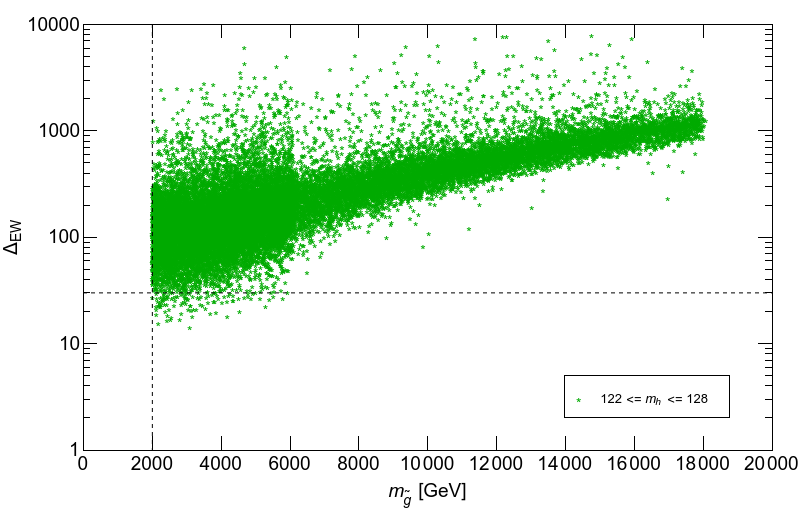}
\caption{Plot of nAMSB parameter scan in the $\Delta_{\rm EW}$ vs. $m_{\tg}$ 
plane.
\label{fig:mgl}}
\end{center}
\end{figure}

In Fig. \ref{fig:mgl}, we show $\Delta_{\rm EW}$ versus $m_{\tg}$.
Here, we see that $m_{\tg}$ ranges from the LHC lower limit of $\sim 2$ TeV up
to $m_{\tg}\sim 6$ TeV before the model becomes unnatural (where
$\Delta_{\rm EW}$ exceeds $\sim 30$). 
The expected range in $m_{\tg}$  will of course have important implications 
for gluino searches at present and planned hadron colliders.
The upper bound $m_{\tg}\alt 6$ TeV is in accord with other SUSY
models: gravity mediation in NUHM2\cite{rns,upper,guts} and in 
mirage mediation\cite{lhc33}. 
The reason is that $m_{\tg}$ feeds into the RG evolution of 
top squark soft terms and a larger value of $m_{\tg}$ therefore 
increases the $\Sigma_u^u(\tst_{1,2})$ values.

\begin{figure}[tbp]
\begin{center}
\includegraphics[height=0.3\textheight]{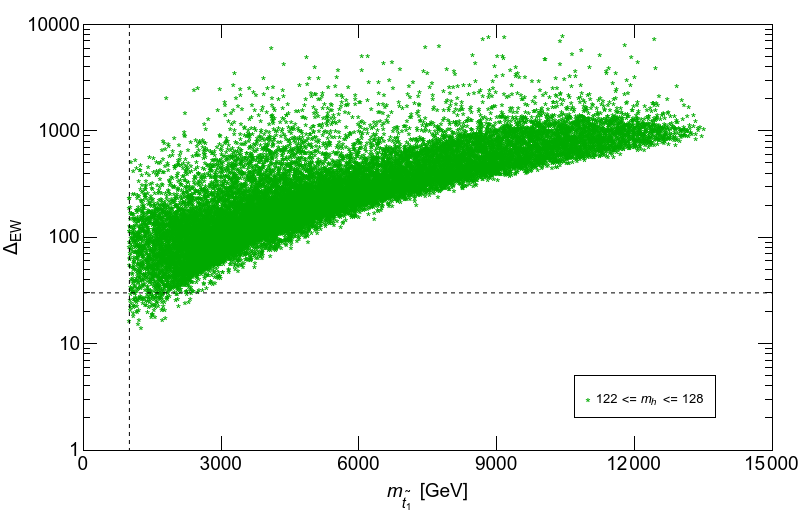}\\
\includegraphics[height=0.3\textheight]{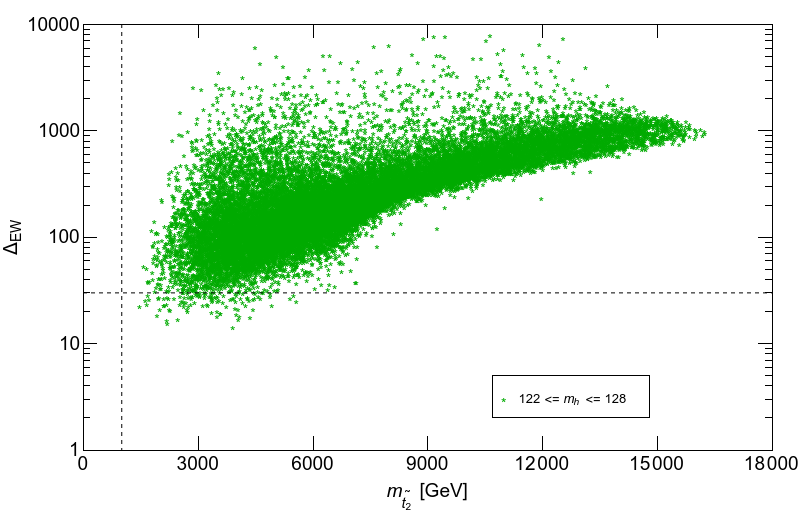}
\caption{Plot of nAMSB parameter scan in the $\Delta_{\rm EW}$ vs. 
{\it a}) $m_{\tst_1}$ and {\it b}) $m_{\tst_2}$ planes. 
\label{fig:mt12}}
\end{center}
\end{figure}

In Fig. \ref{fig:mt12} we show the expected range for top squark masses
$m_{\tst_1}$ (frame {\it a})) and $m_{\tst_2}$ (frame {\it b})).
In frame {\it a}), we see that $m_{\tst_1}$ ranges from its approximate
LHC lower bound of $m_{\tst_1}\agt 1$ TeV up to at most 3 TeV before the 
nAMSB model becomes unnatural. Meanwhile, from frame {\it b}), we see that
$m_{\tst_2}$ can range up to $\sim 6$ TeV. 

\begin{figure}[tbp]
\begin{center}
\includegraphics[height=0.4\textheight]{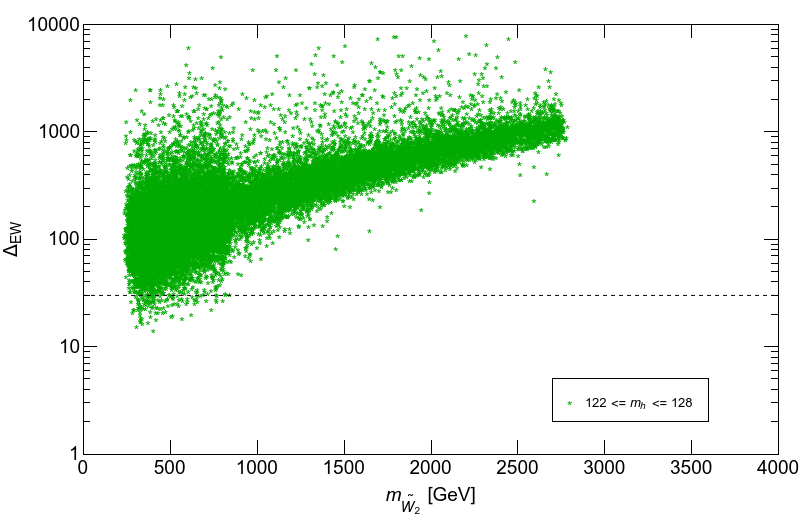}
\caption{Plot of nAMSB parameter scan in the $\Delta_{\rm EW}$ vs. $m_{\tw_2}$ 
plane.
\label{fig:mw2}}
\end{center}
\end{figure}

In Fig. \ref{fig:mw2}, we plot the expected range of wino mass 
$m_{\tw_2}$. In this case, $m_{\tw_2}$ (which is $\simeq m_{\tz_3}$)
ranges from a lower bound $\sim 250$ GeV to an upper bound from naturalness 
of $m_{\tw_2}\sim 800$ GeV. In AMSB models, 
the weak scale wino mass is typically $m(wino)\sim m_{\tg}/8$ so that the
wino mass upper bound arises due to the $m_{\tg}$ limits arising from
$\Sigma_u^u(\tst_{1,2})$. The wino mass range will also have important 
consequences for collider signatures for nAMSB.

\begin{figure}[tbp]
\begin{center}
\includegraphics[height=0.4\textheight]{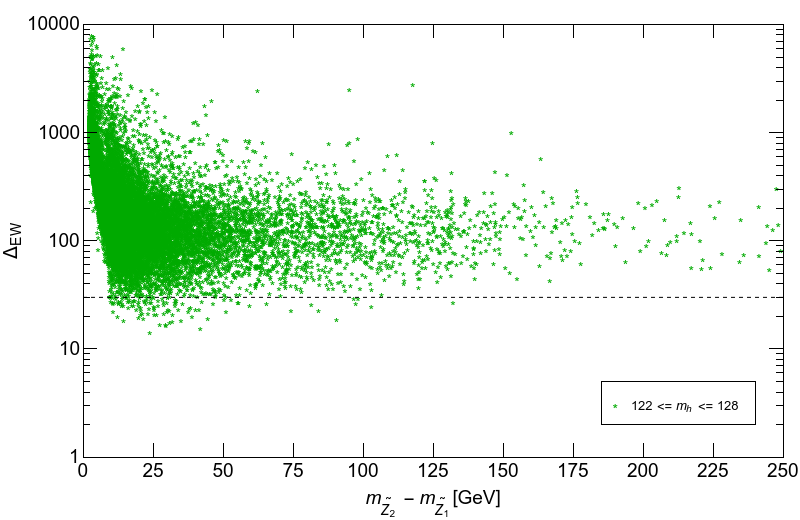}
\caption{Plot of nAMSB parameter scan in the $\Delta_{\rm EW}$ vs. 
$m_{\tz_2}-m_{\tz_1}$ plane.
\label{fig:gap}}
\end{center}
\end{figure}

Lastly, we plot the phenomenologically important mass gap
$m_{\tz_2}-m_{\tz_1}$ versus $\Delta_{\rm EW}$ in Fig. \ref{fig:gap}.
This mass gap enters higgsino pair production signatures at both LHC 
and at linear $e^+e^-$ colliders. Due to the proximity of the winos to the
higgsinos, the mass gap is expected to be larger than in models with unified gaugino masses. Indeed, from the figure we see that $m_{\tz_2}-m_{\tz_1}$
ranges from about 10 GeV all the way up to $100$ GeV.
This may be compared to models with gaugino mass unification where instead 
 $m_{\tz_2}-m_{\tz_1}$ ranges from $\sim 10-25$ GeV typically\cite{rns}.

\section{Consequences for collider and dark matter searches}
\label{sec:lhc_dm}

One of many intriguing aspects of the mAMSB model is that it led to 
rather unique collider signatures-- 
such as the presence of quasi-stable winos in sparticle cascade decays.
In this Section, we will find very different collider signatures for
the nAMSB model.

\subsection{LHC}

\subsubsection{Gluino pair production}

At the CERN LHC, an important SUSY search channel comes from gluino
pair production. In nAMSB, almost always $m_{\tg}>m_{\tst_1}$ so 
that $\tg\to \tst_1\bar{t},\tst_1^*t$ followed by
$\tst_1\to t\tz_{1,2,3}$ or $b\tw_{1,2}$. Thus, gluino pair production events
are expected to be rich in both $t$ and $b$ quarks arising from
gluino cascade decays. Recently, the reach for various LHC luminosity 
upgrades has been estimated for natural SUSY models. It is found in 
Ref. \cite{mgluino} that HL-LHC with $\sim 3$ ab$^{-1}$ of integrated
luminosity has a $5\sigma$ reach in $m_{\tg}$ to about $m_{\tg}\sim 2.8$ TeV.
From Fig. \ref{fig:mgl}, we see that this covers only a small portion of 
nAMSB parameter space. 

Meanwhile, the reach of HE-LHC has also been estimated. 
Using $\sqrt{s}=33$ TeV and 1 ab$^{-1}$ integrated luminosity, 
it is found that HE-LHC reach extends to about 5~TeV\cite{helhc,lhc33}, 
thus covering essentially all of nAMSB parameter space. 
Updated run parameters for HE-LHC have recently been proposed
as $\sqrt{s}=27$ TeV but $L=10-15$ ab$^{-1}$. 
The HE-LHC reach using the lower energy/higher luminosity parameter is likely
comparable to our quoted numbers.

\subsubsection{Top squark pair production}

Top squark pair production $pp\to \tst_1\tst_1^*$ is another important LHC 
search channel.
In nAMSB, we expect $m_{\tst_1}:1-3$ TeV. 
This is to be compared to the $5\sigma$ HL-LHC reach to 
$m_{\tst_1}\sim 1.2$ TeV\cite{hl_atlas_t1}. Thus, in this channel 
again HL-LHC will be able to cover only a small portion of mAMSB parameter 
space. The $5\sigma$ HE-LHC reach extends to $m_{\tst_1}\sim 3.2$ TeV\cite{lhc33}. 
Thus, HE-LHC should be able to cover essentially all mAMSB parameter space via top squark 
pair searches.

\subsubsection{Higgsino pair production} 
From Eq. \ref{eq:mzs}, we find that for $\Delta_{\rm EW}<30$, 
then $m_{\tz_{1,2}},m_{\tw_1}\sim \mu\alt 350$ GeV. 
Thus, higgsino pair production reactions occur at potentially observable 
rates\cite{rns@lhc} at LHC. 
Typically, most of the energy from higgsino pair production
goes into making up the two $\tz_1$ particle's rest mass, so the visible energy release is small, making higgsino pair production reactions challenging to 
see\cite{bbh,mono}.
A way forward has been proposed in References~\cite{lljMET} where one produces
$\tz_1\tz_2$ in association with hard initial state jet radiation.
Then one may trigger on the hard jet (or $\eslt$) and within such events search for low mass, soft  opposite-sign dileptons arising from $\tz_2\to\tz_1\ell^+\ell^-$ decay. Recent search results from CMS have been presented\cite{cms:llj}
and results from Atlas are imminent\cite{atlas:llj}. With HL-LHC, 
this channel may well be able to explore the entire parameter space.
A distinctive feature of the nAMSB model is that the
$\tz_2-\tz_1$ mass gap is expected to be substantially larger than in 
models with gaugino mass unification or in mirage mediation\cite{mirage}
due to the smaller higgsino-wino mass gap.

\subsubsection{Wino pair production}

In SUSY models with light higgsinos, a compelling new signature has emerged\cite{ssdb}:
wino pair production followed by decay to same-sign dibosons (SSdB):
$pp\to\tw_2^\pm\tz_3$ 
with $\tw_2\to W\tz_{1,2}$ and $\tz_3\to W^\pm \tw_1^\mp$.
The higgsinos at the end of the decay chain are again quasi-visible
so one really expects half the time a $W^\pm W^\pm +\eslt$ signal which
has very low SM backgrounds arising mainly from $t\bar{t}W$
and other processes. Signal and background have been estimated in 
Ref's \cite{ssdb,rns@lhc,ssdb2}. It is found that the reach of HL-LHC
extends to about $m_{\tw_2}\alt 1$ TeV. Thus, in this channel as well we
expect HL-LHC to completely cover the nAMSB parameter space. 
If such a signal doesn't emerge at HL-LHC, then the mAMSB model will 
be ruled out. If a signal does emerge, then in Ref. \cite{ssdb2} 
several suggestions have been proposed to extract a measurement of 
the wino masses: via counting, via distributions and via 
$++$ to $--$ charge asymmetry. 
The importance of this channel for the nAMSB model derives from the
expected weak scale gaugino mass ratio in AMSB models 
$M_1:M_2:M_3\sim 0.4:0.13:1$ where winos are expected to be far lighter than 
gluinos (or binos).

\subsection{Linear electron-positron colliders}

Since (simple) natural SUSY models require the presence of light higgsinos
(via Eq. \ref{eq:mzs} and Fig. \ref{fig:param2}{\it b})), then the proposed
International Linear $e^+e^-$ Collider, or ILC, is expected to become
a {\it higgsino factory} for $\sqrt{s}>2m(higgsino)$\cite{ilc}. 
The main production reactions are $e^+e^-\to \tw_1^+\tw_1^-$ and $\tz_1\tz_2$.
In spite of the low energy release expected from these reactions, 
the clean operating environment and low SM backgrounds should allow 
the higgsino pair production events to be easily visible.
These features, along with kinematic restrictions on the events, should allow
for precision mass measurements of $\tw_1$, $\tz_1$ and $\tz_2$.
If ILC is built with extendable energy ranging up to $\sqrt{s}\sim 1$ TeV, then there is a strong chance that direct wino production could also be detected via
the $e^+e^-\to\tw_1^\pm\tw_2^\mp$ channel in nAMSB.

Since the mass gaps $m_{\tw_1}-m_{\tz_1}$ and $m_{\tz_2}-m_{\tz_1}$ depend
sensitively on the higgsino-gaugino mixing, it has been shown in 
Ref's~\cite{ilc,ilcgroup,mext} that the electroweak gaugino 
masses can also be extracted to percent level accuracy. 
Once the EW gaugino masses are known to sufficient accuracy, then they may be run via RGE's to higher energies to test whether or not they unify.
In the case of nAMSB, where $M_2\ll M_1$ is expected, the ILC would 
be able to quickly show that anomaly-mediation is the likely underlying 
SUSY model. 

\subsection{Dark matter: WIMPs and axions}

Dark matter in nAMSB is expected to be a higgsino-like WIMP plus 
SUSY DFSZ axion admixture, as with other natural SUSY models.
As seen in Table \ref{tab:bm}, the $\tz_1$ are thermally underproduced
in the early universe although non-thermal processes such as axino and/or saxion production and 
decay in the early  universe may augment these rates. 
The remaining abundance is expected to be comprised of axions. 
In the case where thermal WIMP production dominates, then indeed the 
bulk of dark matter would be axions. 
Precise estimates of the dark matter abundance require the solution of
eight coupled Boltzmann equations which track the radiation density and number
densities for WIMPs, axions, axinos, thermal- and coherent oscillation- 
production of saxions and gravitino production\cite{bbls}.

To assess WIMP detection prospects, one must account for the diminished abundance of WIMPs 
that is quantified by $\xi\equiv \Omega_{\tz_1}h^2/0.12$ and where
in Table \ref{tab:bm} we would expect $\xi$ as low as $0.075$.
The spin-independent neutralino-proton scattering cross section from 
Isatools is shown in Table \ref{tab:bm}. Mutiplying by $\xi$ and comparing to
recent exclusion limits, it is found the benchmark point to be slightly excluded by recent LUX limits. 
But for the case of the nAMSB model, we expect typically higher
$\sigma^{SI}(\tz_1 p)$ rates because the WIMP-WIMP-$h$ coupling, which enters
the SI detection rate, is a product of gaugino times higgsino component. 
The typically reduced wino mass in nAMSB (as compared to models with gaugino mass unification) raises up the scattering rate somewhat.
Detailed WIMP scattering calculations in this model will be needed for a complete assessment of detectability.
The $\sigma^{SD}$ rate and indirect detection rate (in terms of
$\langle\sigma v\rangle |_{v\to 0}$) are also given. 
Multiplying by $\xi$ and $\xi^2$ respectively, these two rates
are still below current bounds as shown in Ref. \cite{wimp}.

While our benchmark point is nominally excluded, 
even with inclusion of the $\xi$ factor, we remark that 
further entropy dumping in the early universe could 
possibly lower the WIMP abundance even 
further from its thermal value\cite{Bae:2013qr}.
A perhaps more compelling scenario is that the nAMSB model may provide 
a viable niche for light axino dark matter. In usual gravity-mediation, the
axino (and saxion) are expected to gain masses of order 
$\sim m_{3/2}$\cite{cl,kim}. In nAMSB, we would expect the saxion to gain
a bulk soft mass $m_s\sim 1$ TeV but the axino mass could be suppressed leading
to an unstable lightest neutralino which suffers late decay to {\it e.g.}
$\ta +\gamma,Z,h$. In such a case, dark matter would be an axion/axino 
admixture.

Meanwhile, detection of the SUSY DFSZ axion has been shown to be more difficult
than in the non-SUSY models due to the circulation of higgsinos in the
$a-\gamma-\gamma$ triangle coupling\cite{axion}. 
Thus, we do not expect detection of 
the associated axion any  time soon unless the presence of exotic matter in 
the $a-\gamma-\gamma$ coupling leads to an increased axion 
detection rate for microwave cavity experiments.

\section{Concluding Remarks}
\label{sec:conclude}

In this paper, we have proposed a new anomaly-mediation paradigm model
which evades the problems of 1. too low a value of $m_h$, 2. unnaturalness
and 3. winolike LSPs which may be excluded by lack of IDD of dark matter.
Our new model, dubbed natural anomaly mediated SUSY breaking or nAMSB, 
merely incorporates the inclusion of non-universal bulk scalar masses and a
bulk trilinear term $A_0$. The former allows for small $\mu$ as required by 
naturalness and leads instead to a higgsino-like WIMP as LSP. The inclusion of a bulk $A_0$ term allows for large stop mixing which lifts $m_h$ up to 
$\sim 125$ GeV whilst decreasing the top-squark radiative corrections to 
the scalar potential $\Sigma_u^u(\tst_{1,2})$. In fact, these revision were
suggested by the model's creators\cite{RS}.

We computed the sparticle mass spectrum in nAMSB. 
While weak scale gaugino masses are still related as $M_1:M_2:M_3\sim 0.4:0.13:1$ 
leading to wino as the lightest gaugino, the lightest charginos and neutralinos 
are instead mainly higgsino-like (but with a non-negligible wino component).
These modifications bring the model into line with Higgs mass, naturalness 
and dark matter constraints. 
But they also greatly modify the collider and dark matter signatures 
which are expected from anomaly-mediation. 
Instead of quasi-stable charged winos leading to terminating tracks in collider
experiments, now there are more rapidly decaying higgsinos at the bottom 
of the spectra. We computed upper bounds on gluino and top squark masses
in nAMSB and found these to be possibly well beyond reach of HL-LHC
although they should be accessible to HE-LHC. 
However, since higgsinos are required to be not too far from the 100 GeV scale, 
then the $\ell^+\ell^- j+\eslt$ signature should likely be accessible to 
HL-LHC albeit with larger chargino and neutralino mass gaps than in 
models with unified gauginos.
Also, the SSdB signature from wino pair production should be detectable 
over the entire natural range of wino masses in nAMSB leading to a conclusive
test of this model. An ILC operating with $\sqrt{s}>2m(higgsino)$ could 
also discover SUSY and unravel the underlying mediation mechanism via
precision higgsino pair production measurements.

Dark matter is expected to consist of a higgsino-like WIMP plus axion admixture.
Prospects for WIMP detection should be better than in natural models with 
gaugino mass unification due to the presence of rather light winos which enhance
the SI DD scattering rates. Axions may remain difficult to detect.
A further alternative is that the nAMSB model may provide a viable home for
mixed axion/axino dark matter.

\section*{Acknowledgments}

This work was supported in part by the US Department of Energy, Office of High
Energy Physics. 

%

%
\end{document}